\documentstyle[aps,floats,epsfig]{revtex} 

\begin{document}


\title{Ab-initio design of perovskite alloys with predetermined
properties:\\ The case of Pb(Sc$_{0.5}$Nb$_{0.5}$)O$_3$}
 
\author{Jorge \'I\~niguez$^{1,2,*}$ and L. Bellaiche$^1$}

\address{$^1$ Physics Department, University of Arkansas,
Fayetteville, Arkansas 72701, USA\\ $^2$ Departamento de F\'{\i}sica
Aplicada II, Universidad de Pa\'{\i}s Vasco, Apartado 644, 48080
Bilbao, Spain }

\date{\today}

\maketitle

\begin{abstract}
A first-principles derived approach is combined with the inverse Monte
Carlo technique to determine the atomic orderings leading to prefixed
properties in Pb(Sc$_{0.5}$Nb$_{0.5}$)O$_3$ perovskite alloy. We find
that some arrangements between Sc and Nb atoms result in drastic
changes with respect to the disordered material, including ground
states of new symmetries, large enhancement of electromechanical
responses, and considerable shift of the Curie temperature. We discuss
the microscopic mechanisms responsible for these unusual effects.
\end{abstract}




Perovskite alloys with the general form
(A$'$A$''$...)(B$'$B$''$...)O$_3$ are of great technological interest
because of their large piezoelectric and dielectric
responses~\cite{tech}. Examples of their applications are
piezoelectric transducers and actuators, non-volatile ferroelectric
memories, and microelectronic devices. Interestingly, little is known
and understood about the microscopic mechanisms responsible for their
very convenient properties.

One particular aspect that seems very promising both to gain
fundamental knowledge and to produce even more efficient materials is
the effect of atomic ordering on the properties of these
alloys~\cite{Davies,Na00,Hem00,Geo00}. For instance, one may wonder
whether some specific atomic arrangements can generate drastic and
useful changes with respect to the usually-grown disordered
material. Examples of such changes are a large enhancement of the
electromechanical responses of the system or moving its Curie
temperature into a temperature range of interest. If such arrangements
exist, identifying the microscopic origin of their unusual properties
would definitely be of great fundamental and technological
importance. Addressing these issues in an efficient manner requires
(1) a computational technique able to accurately determine the
properties of sufficiently large alloy configurations and providing at
the same time a deep microscopic insight; and (2) the ability to find
the right atomic ordering leading to a given prefixed property,
i.e. to solve the so-called {\sl inverse problem}~\cite{imc}.

The purpose of this Letter is to demonstrate that such an ideal
situation is nowadays possible. More precisely, we have used the
recently-proposed first-principles derived effective-Hamiltonian
method of Ref.~\cite{Bel00b} to tackle the inverse problem of
designing and understanding ferroelectric alloys with prefixed
properties. We focused on Pb(Sc$_{0.5}$Nb$_{0.5}$)O$_3$ (PSN) alloy
because Sc and Nb are chemically very different. (Note that Sc and Nb
belong to different columns of the Periodic Table, which makes PSN an
heterovalent alloy). As a result, we expected that atomic ordering in
PSN might lead to significant changes with respect to the disordered
case. As a matter of fact, we found that some arrangements between the
Sc and Nb atoms (i) greatly enhance the electromechanical responses,
(ii) lead to currently unobserved ground states of orthorhombic and
monoclinic symmetries (while the disordered material adopts a
well-known rhombohedral ground state), and (iii) can considerably
shift the Curie temperature.  Furthermore, the analysis of our
findings suggests that the mechanism responsible for these anomalous
features in PSN can be understood by simple electrostatic
considerations and/or simple phenomenological models. Some promising
directions for future experimental and theoretical research are also
provided.

{\sl Modeling PSN}.-- We first describe the effective-Hamiltonian
approach of PSN. One starts by using the so-called ``virtual crystal
approximation''~\cite{Bel00a} to create a simple system
Pb$\langle$B$\rangle$O$_3$ in which $\langle$B$\rangle$ is a virtual
atom involving a kind of average between Sc and Nb atoms.
The relevant lattice distortions of Pb$\langle$B$\rangle$O$_3$ are
described via a first-principles derived effective-Hamiltonian $H_{\rm
eff}^{\langle{\rm B}\rangle}$ as the one used in Ref.~\cite{ZVR} to
simulate the finite-temperature properties of barium titanate.
This Hamiltonian is then {\sl corrected} to account for the presence
of the two different B atoms in the real alloy. The complete
Hamiltonian has the form:
\begin{equation}
\begin{array}{l}
H_{\rm eff}(\{{\bf u}_i\},\{{\bf v}_i\},\{\eta_l\},\{\sigma_i\}) = \\
  H_{\rm eff}^{\langle{\rm B}\rangle}(\{{\bf u}_i\},\{{\bf
  v}_i\},\{\eta_l\}) + \Delta E (\{{\bf u}_i\},\{{\bf
  v}_i\},\{\sigma_j\}),
\end{array}
\label{eq:heff}
\end{equation}
where ${\bf u}_i$ is the local soft mode in cell $i$, $\{{\bf v}_i\}$
are vectors related to the inhomogeneous strains, and $\{\eta_l\}$ are
the homogeneous strains in Voigt notation. The {\sl soft mode} ${\bf
u}$ is defined as the supercell average of the local modes: ${\bf u} =
\frac{1}{N} \sum_i {\bf u}_i$, where $N$ is the number of cells in the
system~\cite{pol}.  The variables $\{\sigma_j\}$ characterize the
atomic configuration: $\sigma_j = +1$ (resp.  $-1$) indicates there is
a Nb (resp. Sc) atom in cell $j$.

Essentially, the correction term for PSN has the form:
\begin{equation}
\Delta E = \sum_i \sum_{j(i)} \sigma_j Q_{ji} \hat{\bf e}_{ji} \cdot
{\bf u}_i,
\label{eq:deltae}
\end{equation}
where $i$ runs over all the cells and $j$ over the first three
nearest-neighbor shells of cell $i$. Here, $\hat{\bf e}_{ji}$ is the
unit vector joining the B-site $j$ to the B-site $i$; $\{Q_{ji}\}$ are
the alloy-correction coefficients, which only depend on the distance
between sites $i$ and $j$. Interestingly, Eq.~(\ref{eq:deltae}) can be
rewritten as the interactions between the dipole moments $Z^* {\bf
u}_i$ and radial fields ${\cal E}_{ji}$ --- the latter being generated
by the $\sigma_j$, being dependent on $\{Q_{ji}\}$, and acting on the
neighboring cells $i$ ---:
\begin{equation}
\Delta E = - \sum_i Z^* {\bf u}_i \cdot \left[ \sum_{j(i)} {\cal
E}_{ji} \right] = - \sum_i Z^* {\bf u}_i \cdot {\cal E}_i,
\label{eq:cale}
\end{equation}
where ${\cal E}_i$ can be viewed as the total field acting on ${\bf
u}_i$.  We find that the first-principles calculated $Q_{ji}$
parameters all have negative signs. As a result, the field ${\cal
E}_{ji}$ created on site $i$ by a Sc (resp. Nb) atom sitting on a site
$j$ is {\sl attractive} (resp. {\sl repulsive}). This is consistent
with an electrostatic picture of PSN resulting from the fact that
Sc$^{+3}$ (resp. Nb$^{+5}$) atoms are negatively (resp. positively)
charged with respect to the average B-atom valence of $+4$; i.e. to
some extent ${\cal E}_{ji}$ can be viewed as the electric field
created in site $i$ by $\sigma_j$.

Equations~(\ref{eq:heff})-(\ref{eq:cale}) indicate that there is a
competition in PSN between the global tendency of the system towards
the ferroelectric ground state (GS) given by $H_{\rm eff}^{\langle{\rm
B}\rangle}$, which is of rhombohedral symmetry with the local modes
all oriented along the same $\langle 111\rangle$
direction~\cite{Hem00}, and the local tendency of the modes to align
parallel to ${\cal E}_i$, which typically favors an inhomogeneous
local-mode distribution and therefore a decrease of the magnitude of
${\bf u}$.  Recent Monte Carlo simulations using this
effective-Hamiltonian description of PSN have demonstrated its high
accuracy to reproduce direct first-principles results in disordered,
rocksalt-ordered (RS), and other non-trivially ordered
structures~\cite{Hem00,Geo00}. These simulations also reproduce very
well the finite-temperature experimental results for disordered PSN,
once the simulation temperatures are linearly rescaled so that the
calculated transition temperature coincides with the measured
one~\cite{Hem00,resc}. On the other hand, accurately predicting the
properties of samples with local compositions significantly different
from the 50\% average of Sc and Nb may require the use of higher-order
correction terms in Eq.~(\ref{eq:deltae}). Even so, we believe that
all the {\sl trends} discussed in this Letter are qualitatively
correct since Eq.~(\ref{eq:deltae}) includes the first-order (and thus
the most important) correction terms.

{\sl Solving inverse problems}.-- The so-called {\sl inverse Monte
Carlo} is the most convenient computational scheme to search for the
atomic ordering --- i.e. the {\sl solution sample} --- leading to a
prefixed property~\cite{imc}. It is a Monte Carlo annealing done on
the compositional variables $\sigma_j$, and controlled by an energy
functional ${\cal O}[\{\sigma_j\}]$ that is defined so that its
minimum --- i.e. the result of the annealing procedure --- corresponds
to the solution sample we pursue. For instance, one would choose
${\cal O} = -|d_{34}|$ to determine the atomic configuration
exhibiting the highest value of the $d_{34}$ piezoelectric
coefficient. Note that we perform {\sl direct} Monte Carlo simulations
using the effective Hamiltonian of Eq.~(\ref{eq:heff}) to evaluate the
energy functional ${\cal O}[\{\sigma_j\}]$ in every inverse Monte
Carlo step. Such evaluations usually require a long computing time
and, as a result, we typically restricted ourselves to the use of a
$4\times 4\times 4$ supercell (320 atoms or 64 perovskite-like cells),
which is smaller than the $12\times 12\times 12$ supercell of
Refs.~\cite{Geo00,Bel00b} but still much larger than those feasible by
direct first-principles methods.

{\sl Tuning the GS structural properties of PSN}.-- We estimate the GS
structural properties (denoted in the following by the ``0''
superscript) by computing the corresponding thermal-averaged values at
5~K. The inverse problems considered here are listed in
Table~\ref{tab:siminv}. We find that the RS sample maximizes the
magnitude of the GS soft mode, which is consistent with the fact that
the internal field ${\cal E}_i$ is null in any cell $i$ in this sample
and, therefore, no inhomogeneous local-mode distribution resulting
from Eq.~(\ref{eq:cale}) can exist. For all the other inverse problems
indicated in Table~\ref{tab:siminv}, the solution sample can be
described by a single non-null $\sigma_{\bf k}$ coefficient defined
as:
\begin{equation}
\sigma_{\bf k} = \frac{1}{N} \sum_j \exp{(i{\bf k}\cdot{\bf R}_j)}
\sigma_j,
\label{eq:sigmak}
\end{equation}
where ${\bf k}$ is a vector in the first Brillouin zone and ${\bf
R}_j$ is the lattice vector associated to cell $j$~\cite{k}.  Our
results show that the composition modulations described by $k_{\alpha}
= 1/4 \equiv 3/4$ (which correspond to the ``Nb-Nb-Sc-Sc'' sequence
along the $\alpha$ cartesian direction) are related to the decrease of
the $\alpha$ component $u^0_{\alpha}$ of the GS soft mode. This result
is a direct consequence of the electrostatic-related
Eq.~(\ref{eq:cale}), since the other one-dimensional modulations
compatible with the considered $4\times 4\times 4$ supercell ---
i.e. $k_{\alpha} = 0$ (``Nb-Nb-Nb-Nb'' or ``Sc-Sc-Sc-Sc'') and
$k_{\alpha} = 1/2$ (``Nb-Sc-Nb-Sc'') --- result in zero internal field
along the $\alpha$ direction and therefore in no reduction of
$u^0_{\alpha}$. Consequently, the sample defined by a single
$\sigma_{\bf k}$ with ${\bf k} = (1/4, 1/4, 1/4)$ exhibits the
rhombohedral GS with the smallest magnitude of the soft
mode. Similarly, the modulations described by ${\bf k} = (0, 0, 1/4)$
and ${\bf k} = (1/2, 1/2, 1/4)$ present a ``$\Leftarrow \Rightarrow
\Rightarrow \Leftarrow$'' sequence of very strong internal fields
${\cal E}_i$ along the $z$ direction. As a result, the $z$ components
of the local modes align parallel to these fields and the mean value
$u^0_z$ is null. This indicates that PSN samples with an orthorhombic
GS can be produced by conveniently arranging the Sc and Nb atoms. We
also tried to produce a tetragonal GS but only obtained a partial
success, namely a sample characterized by ${\bf k} = (0, 1/4, 1/4)$
with a monoclinic GS for which $|u_x^0| \gg |u_y^0| =
|u_z^0|$. Interestingly, both orthorhombic and monoclinic phases have
never been observed (yet) in PSN materials since the ground states of
the two experimentally-grown samples (disordered and RS) are of
rhombohedral symmetry~\cite{Hem00}.

{\sl Maximum electromechanical responses at low temperature}.-- We now
consider the inverse problem of maximizing the electromechanical
responses of PSN at 5~K~\cite{res}. We focus on the piezoelectric
coefficient $d_{34}$ and the dielectric susceptibility $\chi_{33}$
because we numerically find that they exhibit much larger values than
any other coefficient. Interestingly, we also find that both of them
reach their maximum values for the same atomic ordering.  The solution
sample exhibits $\chi_{33}=8319$ and $d_{34}=4868$~pC/N, to be
compared with the values of 325 and 90~pC/N obtained in disordered
PSN. A striking enhancement of the electromechanical response can thus
be achieved by rearranging the Sc and Nb atoms. (We also obtained
other samples with $\chi_{33} > 4000$ and $d_{34} > 2000$ pC/N, all
sharing the structural features that we discuss in the following.) The
GS of the solution sample is characterized by ${\bf u}^0 = (0.544,
0.567, 0.076)$, where the main feature is the small value of the $z$
component of the soft mode. This reflects the fact that the
$\sigma_{\bf k}$ coefficients that contribute most to the atomic
ordering (around 70\% of the total) have $k_z = 1/4$. Our results thus
suggest that samples with a ``$++--$'' composition modulation in the
$z$ direction, where ``$+$'' (resp. ``$-$'') stands for a plane that
is rich in Nb (resp. Sc), will simultaneously exhibit small values of
$u^0_z$ and large values of $\chi_{33}$ and $d_{34}$ at low
temperatures. However, this modulation cannot be too strong since, for
instance, we numerically found that the atomic configuration
characterized by the single ${\bf k} = (0, 0, 1/4)$ (for which $u^0_z
= 0$, see Table~\ref{tab:siminv}) has small electromechanical
responses.

\begin{figure}[h!]

\centerline{\epsfig{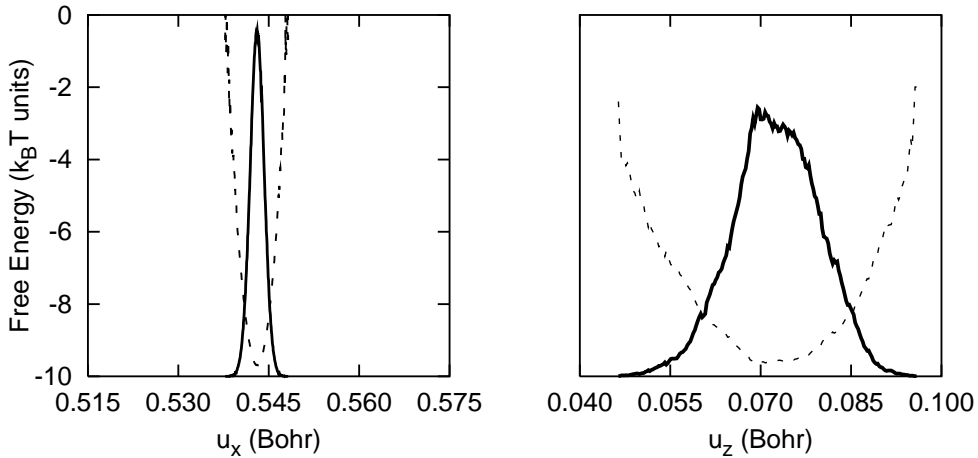}}

\vspace{5mm}

\caption{Probability distribution of $u_x$ and $u_z$ at 5~K (solid
line, in arbitrary units) and the corresponding free-energy map
(dashed line, in $k_B T$ units indicated in the vertical axis) for the
sample with the maximum responses discussed in the text. To obtain these
results, we considered a $12\times 12\times 12$ supercell (trivially
obtained from the $4\times 4\times 4$ solution sample by simple
translation) and performed a $10^6$-sweep Monte Carlo simulation. Only
the region around ${\bf u}^0$ is explored, and the results for $u_y$
are essentially identical to those for $u_x$.}

\label{fig:hist}
\end{figure}

We now discuss the relationship between the small value of $|u^0_z|$
and the large electromechanical responses. Fig.~\ref{fig:hist} shows
the probability distribution of $u_x$ and $u_z$ corresponding to the
Monte Carlo simulation at 5~K of the solution sample, as well as the
free-energy map of the system around ${\bf u}^0$. This latter is
calculated out of the probability distribution as done in
Ref.~\cite{Rad95}. The two main features of Fig.~\ref{fig:hist} are
(1) the {\it broadness} of the $u_z$ distribution, which is
responsible for the large value of $\chi_{33}$ (see Ref.~\cite{res})
and (2) the {\it flatness} of the corresponding free-energy well,
which indicates the ease of rotating the polarization around
$u^0_z$. Interestingly, both features are typically associated with a
small value of $|u^0_z|$ within the usual Devonshire-Landau
description of ferroelectric perovskites~\cite{dev}.  Moreover, we
numerically find that $u_z$ and $\eta_4$ are strongly coupled in our
effective Hamiltonian for PSN, which implies that their thermal
fluctuations are very correlated. Consequently, according to
Ref.~\cite{res}, a large value of $\chi_{33}$ typically corresponds to
a large value of $d_{34}$, as we indeed find.

\begin{figure}[h!]

\centerline{\epsfig{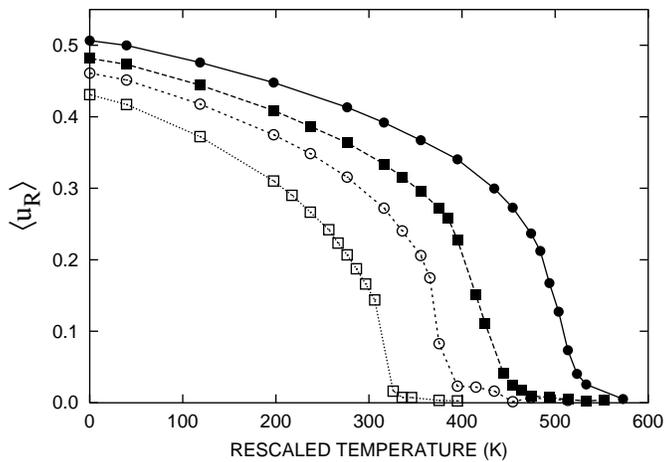}}

\vspace{5mm}

\caption{Temperature evolution of $\langle u_{\rm R} \rangle$ (see
caption of Table~\ref{tab:siminv}) for four PSN samples. These samples
all undergo a single (cubic to rhombohedral) phase transition, and are
chosen so as to exhibit different values of the GS soft mode. The
values of $u^0_{\rm R}$ considered here are: $0.507$~Bohr (RS, which
exhibits the largest possible value of $u^0_{\rm R}$), $0.484$~Bohr (a
intermediate sample generated by solving the appropriate inverse
problem), $0.461$~Bohr (disordered sample), and $0.431$~Bohr (the
sample defined by ${\bf k} = (1/4, 1/4, x1/4)$, which exhibits the
lowest possible value of $u^0_{\rm R}$). All the results correspond to
the $2\times 10^5$-sweep Monte Carlo simulation of a $12\times
12\times 12$ supercell, which thus guarantees a good convergence in
the calculated transition temperatures. The temperatures have been
rescaled so that the calculated $T_c$ of the disordered sample (944~K)
coincides with the experimental value (373~K).}

\label{fig:cont} 
\end{figure}

{\sl Tuning the finite-temperature properties of PSN}.-- Studying PSN
at finite (relatively-high) temperatures requires long Monte Carlo
simulations in order to obtain good statistics. Thus, due to
computational limitations we decided not to {\it directly} solve
finite-temperature (FT) inverse problems. However, some links between
the GS and FT properties are expected on the basis of well-known
models that share some essential features with our
effective-Hamiltonian description of PSN. More precisely, the
so-called $\phi^4$ model~\cite{Bru80} suggests the following
relationship: the larger the value of $|u^0|$, the deeper the
corresponding free-energy well at 0~K and, therefore, the higher the
Curie temperature $T_c$. Hence, we thought it was pertinent to check
whether the Curie temperature of PSN could be tuned by changing the
value of the GS soft mode. Consequently, we studied the FT behavior of
four samples that all undergo a single cubic to rhombohedral
transition but exhibit different values for their GS soft modes. Our
results, shown in Fig.~\ref{fig:cont}, confirm that it is indeed
possible to move the Curie temperature to a temperature range of
interest by designing alloys with a prefixed value of $|u^0|$.  It is
remarkable that relatively-small changes in the GS soft mode result in
such large variations in $T_c$. Interestingly, the transition of the
sample with the smallest $|u^0|$ takes place near room temperature. It
is important to notice that one can tune the FT electromechanical
responses of the system by moving $T_c$, because these responses reach
their maximum values around the paraelectric to ferroelectric
transition.

In summary, we have implemented a method to solve inverse problems in
perovskite alloys by using the first-principles effective-Hamiltonian
approach of Ref.~\cite{Bel00b}, and applied it to PSN. We predict that
the system properties (GS symmetry, electromechanical responses, Curie
temperature, etc.) can be greatly modified by varying the atomic
ordering. Furthermore, all the anomalous features we predict can be
essentially understood by simple electrostatic considerations and/or
simple phenomenological models. As a result, (1) we have indicated
which kinds of alloy structures may lead to currently-unobserved
structural phases as well as to giant dielectric and piezoelectric
responses, and (2) we expect similar results to occur in other
heterovalent perovskite alloys. Finally, we hope that our predictions
will stimulate attempts at experimental growth of such samples.

This work is supported by Office of Naval Research Grants
No. N00014-00-1-0542 and No. N00014-01-1-0600, National Science
Foundation Grant DMR-9983678 and Arkansas Science and Technology
Authority Grant N99-B-21.  We thank Aaron George, Alberto Garc\'{\i}a,
and J. M.  P\'erez-Mato for comments and discussions. J. I. further
thanks the financial support of the Government of the Basque Country
and the hospitality of all the members of the Physics Department of
the University of Arkansas.

\begin{table}[h!]

\caption{Inverse problems related to several GS properties of
PSN. First column: energy functional ${\cal O}$. The ``R'' subscript
indicates a rhombohedral symmetry (i.e. ${\bf u}_{\rm R} = u_{\rm
R}(1, 1, 1)$). Second column: solution configurations. ``RS'' stands
for the rocksalt structure while we indicate the ${\bf k}$ point
leading to the non-null $\sigma_k$ coefficient (see text) for the
other cases. Third column: GS soft mode ${\bf u}^0$. In one case, we
indicate two ``good'' solutions. We include the information about the
disordered sample for comparison.}

\begin{center}
\begin{tabular}{rcc}

${\cal O}$ & sol. conf. & ${\bf u}^0$ (Bohr) \\ \hline

$-|{\bf u}^0_{\rm R}|^2$ & RS & $(0.507,0.507,0.507)$\\

$|{\bf u}^0_{\rm R}|^2$ & $(1/4,1/4,1/4)$ & $(0.431,0.431,0.431)$\\

$(u_z^0)^2$ & $(0,0,1/4)$ & $(0.552,0.552,0.000)$ \\

& $(1/2,1/2,1/4)$ & $(0.484,0.484,0.000)$\\ 

$(u_y^0)^2+(u_z^0)^2$ & $(0,1/4,1/4)$ & $(0.665,0.242,0.242)$\\ \hline

& disordered & $(0.461,0.461,0.461)$

\end{tabular}
\end{center}

\label{tab:siminv}
\end{table}

\begin{thebibliography}{99}
\bibitem{pol} Note that the polarization of the simulated system is given by
$(Z^*/v) {\bf u}$, where $Z^*$ is the Born effective charge associated
to the local soft modes and $v$ is the unit cell volume.
\bibitem{k} We omit the $2\pi/a$ factors of the wave vectors, where
$a$ is the lattice parameter of the perovskite-like cell.
\end{thebibliography}
\end{document}